\documentclass[sigconf, nonacm]{acmart}

\AtBeginDocument{%
  \providecommand\BibTeX{{%
    \normalfont B\kern-0.5em{\scshape i\kern-0.25em b}\kern-0.8em\TeX}}}

\setcopyright{acmlicensed}
\copyrightyear{2023}
\acmYear{2023}
\acmDOI{XXXXXXX.XXXXXXX}

\acmConference[Conference acronym 'XX]{Make sure to enter the correct conference title from your rights confirmation emai}{Time}{Place}
\acmISBN{ACM ISBN}

\usepackage{listings}
\usepackage{dirtytalk}
\usepackage{graphicx}
\begin{document}

\title{"I don't like things where I do not have control": Participants' Experience of Trustworthy Interaction with Autonomous Vehicles}

\author{Ana Tanevska}
\email{ana.tanevska@it.uu.se}
\orcid{0000-0002-2628-4123}
\affiliation{%
  \institution{Department of Information Technology, Uppsala University}
  \city{Uppsala}
  \country{Sweden}
}

\author{Katie Winkle}
\email{katie.winkle@it.uu.se}
\orcid{0000-0002-3309-3552}
\affiliation{%
  \institution{Department of Information Technology, Uppsala University}
  \city{Uppsala}
  \country{Sweden}
}

\author{Ginevra Castellano}
\email{ginevra.castellano@it.uu.se}
\orcid{0000-0002-2841-6791}
\affiliation{%
  \institution{Department of Information Technology, Uppsala University}
  \city{Uppsala}
  \country{Sweden}
}

\renewcommand{\shortauthors}{Tanevska et al.}

\begin{abstract}


With the rapid advancement of autonomous vehicle (AV) technology, AVs are progressively seen as interactive agents with some level of autonomy, as well as some context-dependent social features. 

This introduces new challenges and questions, already relevant in other areas of human-robot interaction (HRI) - namely, if an AV is perceived as a social agent by the human with whom it is interacting, how are the various facets of its design and behaviour impacting its human partner? And how can we foster a successful human-agent interaction (HAI) between the AV and the human, maximizing the human's comfort, acceptance, and trust in the AV?

In this work, we attempt to understand the various factors that could influence naïve participants' acceptance and trust when interacting with an AV in the role of a driver. Through a large-scale online study, we investigate the effect of the AV's autonomy on the human driver, as well as explore which parameters of the interaction have the highest impact on the user's sense of trust in the AV. Finally, we analyze our preliminary findings from the user study within existing guidelines on Trustworthy HAI/HRI.

\end{abstract}






\maketitle

\section{Introduction}

As their technology progresses, autonomous vehicles (AVs) evolve from augmented technology to increasingly more complex agents, required to collaborate with humans as well as with other artificial agents in a multi-agent system \cite{ferber1999multi}. AVs become more autonomous in their navigation, perception, and decision-making, which necessitates an ability to perceive and understand the other agents' intentions and states, to communicate their own intentions, and coordinate their actions \cite{beydoun2009security, tanevska2023communicating, malik2021collaborative, lee2021road}.

However, while a successful interaction with other artificial agents primarily requires an understanding of the other agent's actions and intent, introducing humans into the equation also brings up other relevant topics including safety, transparency, predictability, and trust \cite{eu_ethics_guidelines_2019, christ2016human, okamura2020calibrating}. 

To create an efficient and safe human-AV interaction that the users will perceive as comfortable and trustworthy, we need to research the impact of the AV's interface and behavior on naïve participants' experience with it, and understand how their trust and acceptance is affected \cite{detjen2021increase, raats2020trusting, sripada2021automated,rettenmaier2021matter}. Additionally, since the development of social artificial agents with a degree of autonomy and awareness requires a more in-depth understanding of how the agent's \textit{autonomy} impacts the human, we need to include also this factor when planning how to build a trustworthy and ethical human-agent interaction (HAI) \cite{khavas2020modeling,gao2019fast}.

In our research, we tackle this issue by conducting user-centered studies \cite{winkle2021leador} with the potential end users of the AVs, in which we discuss different interaction scenarios, analyze the information the AV is processing and using in its decision-making, and develop the users' requirements for a trustworthy interaction. We began by conducting a small participatory design study \cite{tanevska2023communicating} consisting of two different driver-AV interaction scenarios, where we worked together with the participants to develop different interfaces for the AV for their preferred levels of transparency, as well as analyzed how comfortable they were with the different levels of autonomy (LoA) in which the AV functioned.

We follow the participatory workshop with a larger-scale online study, whose design was informed by the findings of our workshop. With it, we wish to: 1) further investigate how the AV's transparency interacts with the users' sense of agency, 2) analyze any correlations between participants' demographics and experiences and their perception of the AV, and 3) ground our findings following a defined set of design guidelines for Trustworthy HAI.

The rest of the paper is organized as follows: Section 2 presents the design guidelines for trustworthy interaction we are using in our evaluation and analysis; Section 3 goes over the experimental materials and methods; Section 4 presents our preliminary results, and Section 5 concludes the paper, outlining our next steps.

\section{Design guidelines for trustworthy HRI and HAI}\label{sec:first_2}

To explore the concepts of user agency, oversight, and transparency in our studies, we opted to ground our analysis using an existing set of design guidelines based on the EU Trustworthy AI concepts \cite{eu_ethics_guidelines_2019}, which were created specifically for investigating trust and acceptance in HRI and HAI scenarios \cite{calvobarajas2024guidelines}. The guidelines are:

\begin{itemize}
    \item \textbf{Security and Privacy}: Comply with data protection regulations and promote transparent documentation.  Implement data minimization and anonymization.

     \item \textbf{Trust and Transparency}: Foster transparent communication.  Design for adaptability, feedback processing, and environmental responsiveness. Incorporate the appropriate level of human likeness or anthropomorphism. 

     \item \textbf{Safety and Performance}: Ensure the physical and emotional safety of users.  Incorporate social adaptability during the interaction.
     
     \item \textbf{Competency and Control}: Design for task management and task integration in sociotechnical systems. Incorporate context-aware autonomy levels and flexibility.
     
     \item \textbf{Positive Experience}: Ensure respect for human rights, diversity, and inclusion. Design for long-term use and novelty effect mitigation. 
\end{itemize}








\section{Understanding user preferences through participatory design}

Prior to embarking on the larger-scale online study on the concepts of agency, oversight and transparency, we conducted a smaller participatory design study in the form of a workshop \cite{tanevska2023communicating}.
The aim of the participatory workshop was to investigate how the AV's autonomy may impact the human users, and which facets of the interaction have the highest impact on the user's sense of trust in it. This was represented with the following research questions: 

- \textit{To what extent should the AV act autonomously? }

- \textit{How can human users be in the loop and monitor the interaction?}

\begin{figure}[h]
    \centering
    \includegraphics[width=0.75\columnwidth]{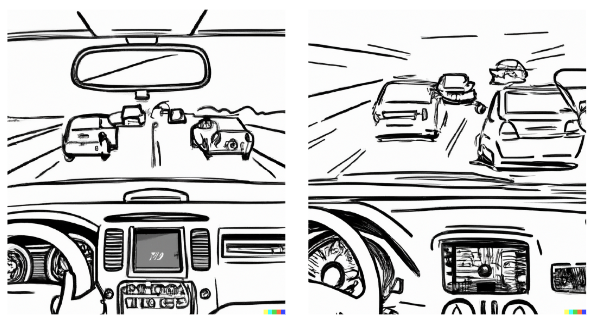}
    \caption{\textit{Highway} driver-AV scenario. AI-generated images of a car's dashboard with the driver's POV, car driving on a busy highway and another car cutting in from the right.}
    \label{fig:scenario1}
\end{figure}

Following the demographics questionnaires, participants took part of the interactive workshop, comprising two use case scenarios (presented to users as series of computer-generated images\footnote{Images generated with https://openai.com/dall-e-2} on a flipchart). In both of them, the participants had to imagine themselves in the role of the driver of the AV (i.e. as sitting in the driver seat).
In the first scenario (Fig. \ref{fig:scenario1}), the AV was on a busy highway, driving in the middle lane, with an unpredictable moment when another car merged from the right side at an unsafe distance and speed, risking impending collision.

\begin{figure}[h]
    \centering
    \includegraphics[width=0.75\columnwidth]{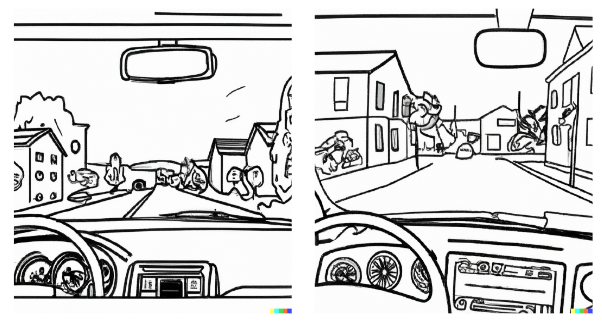}
    \caption{\textit{Suburbs} driver-AV scenario. AI-generated images of a car's dashboard with the driver's POV, car driving on an empty neighborhood street, with a school building coming up on the left.}
    \label{fig:scenario2}
\end{figure}

In the second scenario (Fig. \ref{fig:scenario2}), the AV was on an unfamiliar empty suburban street, when it suddenly slowed down to 20 km/h without any visible obstacle in view. As the scenario progressed, a school that was previously obscured behind trees came into view.

\begin{figure}[h]
    \centering
    \includegraphics[width=0.9\columnwidth]{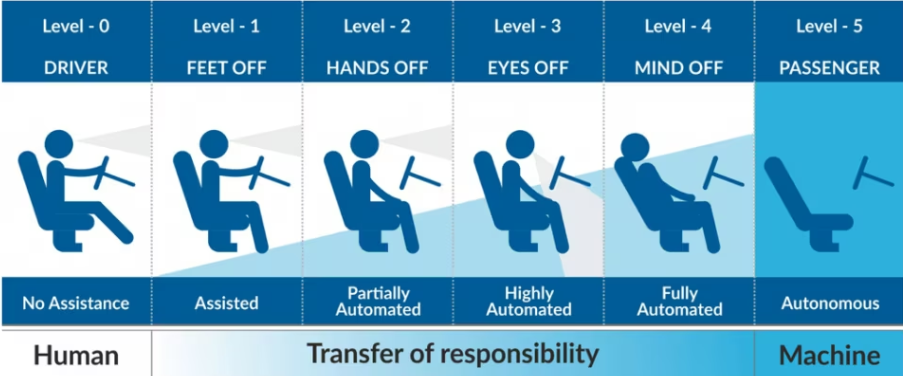}
    \caption{The six SAE Levels of Autonomy. In our study we worked with Levels 0 to 3, i.e. "Driver", "Feet off", "Hands off", and "Eyes off".}
    \label{fig:sae}
\end{figure}

Participants took part in two activities - modulating the \textbf{Level of Autonomy (LoA)}, and designing a \textbf{user interface} for interaction between the car and the driver. The levels of autonomy we worked with in this study were the SAE's Levels of Autonomy \cite{sae2018taxonomy}, specifically Levels 0 through 3 (see Fig.\ref{fig:sae}). For each scene in both scenarios, participants were asked to select the highest level of autonomy they were comfortable with.

The second workshop activity utilized the information that the AV uses to calculate its SA and select its next action. For each scenario participants were asked to work together as a group to design the interface for the AV (using post-it notes and drawing on the flipchart images), deciding on the amount and type of information they wanted shown. The designed interface was then considered through all scenes of each scenario, and any eventual modifications were made during the scene changes.

Two main designs for the interface resulted from the workshop: a minimal, unobtrusive interface similar to modern car GPS navigators; and a much more complex design containing visual depiction of all of the information the AV is processing and its upcoming actions. Participants who opted for the first design tended to want explicit communication from the AV only in moments of emergencies (e.g. \say{another car is about to collide with me}), whereas the proponents of the second design wished to be able at any moment to see how the AV is processing and understanding the scene, and how it selects its actions.

In our previous publication we discussed in-depth the details of the participatory design study and its findings \cite{tanevska2023communicating}. The main relevance of the workshop results for this work are the two distinct interfaces as individuated by the participants, which informed the design of our online study and its two information levels - with the minimal interface giving us the Low Information condition, and the complete interface the High Information one.

\section{Materials and Methods}

\subsection{Participants}

In total 206 participants took part in our study, equally distributed between male and female\footnote{Legal sex as reported on their current documents. The self-reported gender ratio instead was 104:101:1 (F:M:X)}, and their age ranged from 20 to 73 years (M=36.42, SD=13.185). Participants were recruited via the crowd-sourcing platform Prolific\footnote{https://www.prolific.com/}. We used the platform's filters to recruit English-speaking participants with a valid driver's license. 

From the recruited participants, 29.1\% had been active drivers for over 20 years, 18.9\% between 11 and 20 years, 42.2\% between 3 and 10 years, and 9.7\% had been driving less than 2 years.
Regarding the frequency of driving, 66\% of participants drove daily, 28.6\% drove several times a week, and the rest were occasional drivers, with only 2 participants no longer being active drivers.

\subsection{Experimental Design and Protocol}
Our study followed a 2×2 between-subjects design, corresponding to the designed information level (High Information vs. Low Information) and the order in which participants experienced the two scenarios (Highway-first vs. Suburbs-first). Participants were randomly assigned to one of the four groups.

At the start, participants filled one questionnaire on demographics (including information regarding their driver experience), and one on their driver behavior (an adapted and abridged version of the DBQ \cite{reason1990errors}). 
This was followed by a brief presentation on AVs which introduced participants to the term, outlined the types of information the AV processes when making its decisions, and presented participants with the concepts on autonomy in AVs and how it's quantified (also here we used the SAE's Levels of Autonomy \cite{sae2018taxonomy}, see Fig.\ref{fig:sae}). It also broached the concept of social interaction with AVs, and what it entails.

After the presentation, there was a manipulation check (describing one SAE level and asking participants to select which one it is), after which participants filled the last questionnaire, which was an adaptation of the NARS \cite{nomura2004psychology}, aimed at identifying and quantifying any negative attitudes or preconceptions participants may have towards AVs. Our version (referred to as AV-NARS) had 10 items, adapted to fit the context of interaction with AVs, e.g. "I would feel nervous driving an autonomous car while other people could observe me." instead of the original "I would feel nervous operating a robot in front of other people.". 

After the questionnaires, participants were presented with the same two use case scenarios (the Highway and Suburbs ones) from the participatory design workshop as described in Section 3. Each of the two scenarios consisted of three scenes, with the second scene in both of them presenting the unexpected moment (the second car merging unsafely, or the AV slowing down in the empty street).

\begin{figure}[t]
    \centering
    \includegraphics[width=\columnwidth]{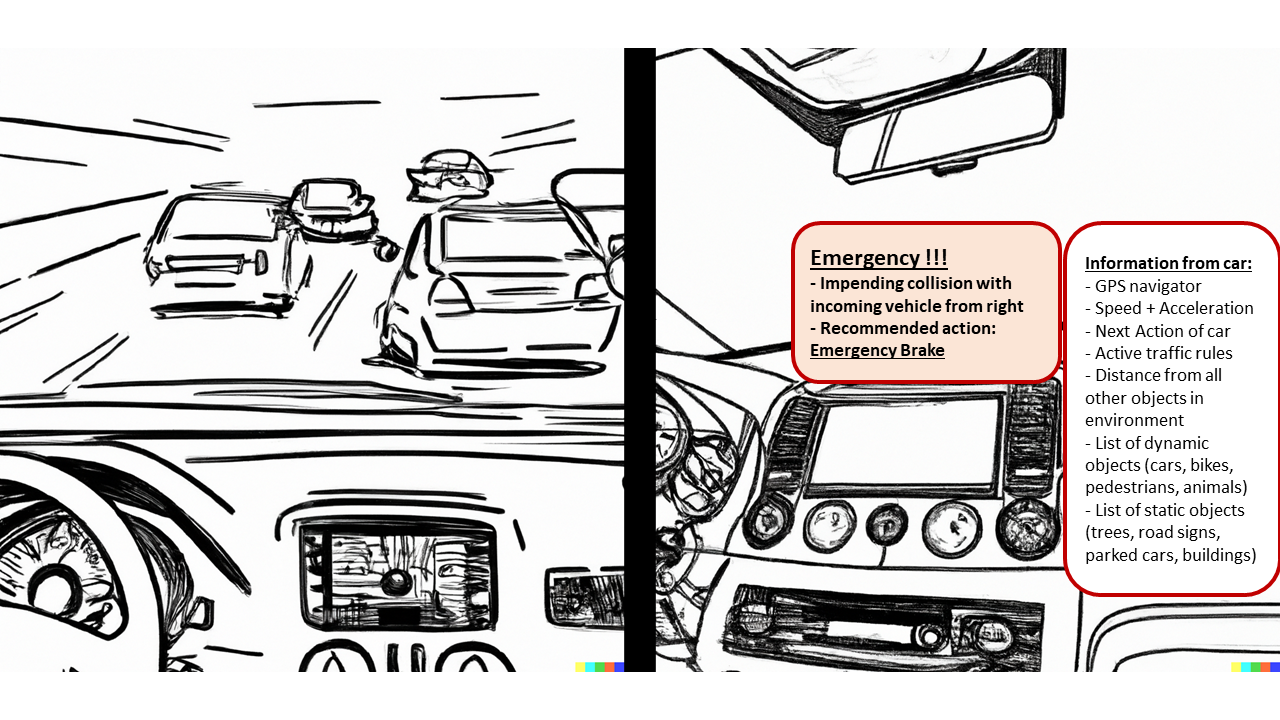}
    \caption{\textit{Highway} driver-AV scenario, plus a High Information interface for the AV. AI-generated images of a car's dashboard with the driver's POV, car driving on a busy highway and another car cutting in from the right.}
    \label{fig:scenarioFull}
\end{figure}

Each of the three scenes consisted of a computer-generated image of the dashboard view of the scene plus an overview of the information presented to them (see Figure \ref{fig:scenarioFull}), and a roleplay-like narration of what happens in the scene. Participants were then asked "What do you do?" and could select from a multiple choice answers, where each choice corresponding to a level of autonomy between 0 and 3. E.g. for the first scene of Scenario 1, the possible answers were: \{\textit{"I do nothing, I let the car continue navigating", "I quickly check my phone for messages", "I check out the area around me and look around for anything interesting", "I focus my eyes on the road and check for any people or animals that may approach the road", "I put my hands back on the wheel", "I take over the driving control from the car"}\}
 
After their action selection, participants answered questions on their confidence and comfort levels with the situation, and their trust in the AV. The questions are described in greater detail in the following section.

\subsection{Data collection}

We grouped the collected data from the survey in four categories:

\emph{- Questionnaires.}
Participants answered questionnaires on demographics, adapted DBQ (driver behavior questionnaire) and adapted AV-NARS (negative attitudes towards AVs).

\emph{- Action choice.}
The selection of action following each of the six scenes. A multiple-choice question where each choice corresponded to a level of autonomy between 0 and 3. This was averaged in the cases where participants chose more than one option, so that also the final LoA score ranged from 0 to 3.

\emph{- Confidence and comfort.}
After each action decision, participants were asked to rate how confident they felt about their decision (5-point Likert scale with descriptive ranking, ranging from "Not confident at all" to "Completely confident"). Following that, they were also asked about their comfort with the situation, which was a single-choice question where they could pick "relaxed, calm", "anxious, tense", "neutral", and "unsure". The comfort score was rated as -1, 0, or 1.

\emph{- Trust in the AV.}
For their final evaluation, participants were asked about their trust in the car via a multiple choice question, with the options being: \{\textit{"I feel that the car is reliable", "I can trust the car's actions to be safe", "I don't feel safe in the car", "I can rely on the car to make safe decisions for my wellbeing", "I feel I can follow the car's instructions", "I don't trust the behaviour of the car"}\}. The answers were coded as 1 or -1 depending on whether they expressed a positive or negative trust sentiment, and then summed up to form the total Trust score, which ranged from -2 to 4. In addition, if participants felt that none of the statements were accurate about the current situation and their decision, they had an empty text field to provide free answers.

\section{Results}

Although we have yet to conduct a full statistical analysis on the survey data, we performed a preliminary Pearson Correlation Coefficient calculation for the results of the DBQ and AV-NARS questionnaires, and the participant ratings of the Levels of Autonomy (LoA) and Trust in the AV.

Our analysis showed no correlation between people's DBQ and AV-NARS scores ($r$(204)=-0.075, $p$=0.286), nor between their DBQ score and their LoA ($r$(204)=-0.007, $p$=0.915) or Trust ($r$(204)=0.120, $p$=0.087) ratings. In other words, participants' driving style (regarding their errors while driving and risk-taking behaviours) did not impact their attitudes towards AVs.

We did however note a weak-moderate negative correlation between the AV-NARS scores and the LoA ratings ($r$(204)=-0.369, $p$<0.001), and a moderate negative correlation between the AV-NARS score and the Trust ratings ($r$(204)=-0.583, $p$<0.001). This is in line with the findings from our earlier participatory design study, which also showed that the less negative attitudes towards AVs users may have, the higher their trust in the AV and the level of autonomy they’d be comfortable with the AV possessing.

Additionally, looking into the correlations between the average LoA and Trust scores, as well as the separate Scenario 1 and Scenario 2 ratings of the same, we found a general positive correlation between the Trust and LoA ratings ($r$(204)=0.576, $p$<0.001), both for the average ratings of the entire study, and for the separate scenario ratings (meaning the pairs of Trust/LoA for Scenario 1 and Scenario 2). Looking at the two ratings individually instead, we found a positive correlation between the Trust rating of Scenario 1 and Scenario 2 ($r$(204)=0.580, $p$<0.001), but no correlation between the same ratings for the LoA ($r$(204)=-0.002, $p$=0.980). This tells us that while participants’ trust and level of autonomy tend to be correlated, their trust towards AVs seems to be more intrinsic and stable across evaluations, whereas the level of autonomy depends heavily on the scenario.  

Our last analysis looked into the free answers provided as alternative to the Trust question. In total 14 participants provided additional text responses, most of them after the very first Scene of the experiment (for some the Highway, for others the Suburbs), but a couple of participants answered after each scene as well. We group the answers considering the categories from Section 2:

\begin{itemize}
        
     \item \textbf{Safety and Performance}: Of note in this category was that the users' main concerns were regarding the situation and behaviour of \textit{other} drivers in the scene, and not so much regarding the car's abilities:
        \begin{itemize}
            \item I don't like things where I do not have control (All Scenes). 
            \item Reliable but not enough to cross over all those lanes of traffic for me. Too many risky driver behaviour (Scene 1).
            \item My anxiousness would be about what could influence the car from outside and not the car itself (Scene 1).
            \item In a situation like that, I take control of the car and drive it myself (Scene 2).
            \item I feel safe, but i prefer to take over to handle the situation by myself (Scene 3).
            \item I want to make sure that my safety along with anyone else that is in the school zone is a priority (Scene 3).
        \end{itemize}

     \item \textbf{Trust and Transparency}: We sort here the replies from users which expressed their trust as a function of the AV's ability to communicate its behaviour, highlighting the value of transparent communication from the car and a responsive behaviour:
        \begin{itemize}
        \item Although the car is doing what its supposed to do, I would be happier that I can take over if necessary (Scene 1).
        \item I would have confidence in the car but would still be aware of any problems that may arise (Scene 1).
        \item I can moderately trust the car's actions (Scene 1).
        \end{itemize}
        
     \item \textbf{Competency and Control}: In this category we consider the replies from users who focused on the AV's capabilities when evaluating their trust in its decisions:
     
    \begin{itemize}
        \item I am willing to trust the car, but will remain cautious should anything go wrong I am at the ready to take over and correct mistakes (Scene 1). \\Again, I am willing to test the cars capabilities, however I am focused and at the ready to correct any mistakes that may occur (Scene 2).\\ I have not tested the car enough to fully trust it, so I will take over to avoid collision. Once tested and I can trust the car fully i will let it do its thing (Scene 3).
        \item I will be cautious and wait to see how the car responds to this (Scene 2).
            \item ((after taking over control)) 20 km/h is too slow (Scene 3).
        \item I'm worried that the car won't respond safely because the sign was blocked (Scene 3).
        \end{itemize}

    \item \textbf{Security and Privacy}: None of the free text replies indicated any security or privacy concerns from users.

     \item \textbf{Positive Experience}: None of the free text replies indicated any concerns regarding potential disrespect of users' rights or lack of inclusion.

\end{itemize}







\section{Conclusion}
This short paper presents our preliminary findings from a large-scale online study on the factors of human-AV interaction which interact with the concepts of agency, oversight and transparency.

We were interested in investigating how different levels of transparency on the AV's side can influence the user's comfort and trust in the AV, as well as the maximum level of autonomy they would be comfortable with the AV possessing.

From our first analysis, we discovered that the type of scenario (Highway/Suburbs) in which the users are interacting with the AV tends to be the most important factor when selecting a level of autonomy. Trust, on the other hand, was influenced both by the interaction scenario as well as the amount of information that was presented on the AV's interface.

We also looked into the qualitative data collected from a portion of the users, where the most common concepts that stood out were the safety and performance of the AV, the transparency of the AV and its correlation with trust, and the AV's competency and control. From the original set of design guidelines \cite{calvobarajas2024guidelines}, the two concepts that were not brought up explicitly by any user were the security and privacy, and the positive experience. We would like to further investigate in follow-up studies in which contexts these two concepts may become more relevant to users (their importance when \textit{designing} HAI scenarios is absolute and beyond questioning).

In our next steps, we plan to continue with more qualitative and statistical analysis of the survey data to further explore the correlations between the AV's design and behaviour, and people's feelings of confidence, comfort, and trust. We will also analyze further the open-ended answers of the participants, and specifically how their valence (more positive or negative) correlates with the other metrics in the study (the LoA selected by the participant, their comfort and trust scores, etc). 

We aspire to use our findings for the eventual creation of a computational model of a human driver, which would then be integrated in our multi-agent system \cite{tanevska2023communicating, calvagna2023using}. We view this integration as having the potential to improve the overall performance and effectiveness of the multi-agent system by leveraging the cognitive capabilities of both humans and machines, as well as contribute to more natural, trustworthy, and long-term interaction between humans and artificial agents.

\begin{acks}
This work is supported by the Horizon Europe EIC project SymAware (\url{https://symaware.eu}) under the grant agreement 101070802. 
\end{acks}

\bibliographystyle{ACM-Reference-Format}
\bibliography{sample-base}

\end{document}